\documentclass[conference]{IEEEtran}
\IEEEoverridecommandlockouts
\usepackage{filecontents}
\usepackage{cite}
\usepackage{subfig}
\usepackage[normalem]{ulem}
\usepackage[T1]{fontenc}
\usepackage{amsmath,amssymb,amsfonts}
\usepackage{algorithmic}
\usepackage{graphicx}
\usepackage{textcomp}
\usepackage{xcolor}
\def\BibTeX{{\rm B\kern-.05em{\sc i\kern-.025em b}\kern-.08em
    T\kern-.1667em\lower.7ex\hbox{E}\kern-.125emX}}

\begin{document}

\title{ A Study of Tourist Sequential Activity Pattern through Location Based Social Network (LBSN)
}
\author{\IEEEauthorblockN{Anmoila Talpur}
\IEEEauthorblockA{\textit{Institute for Sustainable Industries \& Liveable Cities,} \\
\textit{VU Research, Victoria University}\\
 Australia \\
anmoila.talpur@live.vu.edu.au}
\and
\IEEEauthorblockN{Yanchun Zhang}
\IEEEauthorblockA{\textit{Institute for Sustainable Industries \& Liveable Cities,} \\
\textit{VU Research,Victoria University}\\
 Australia \\
yanchun.zhang@vu.edu.au}
}

\maketitle

\begin{abstract}
Sequential Pattern Mining (SPM) is an important component in establishing patterns and mining trends of certain activities. In the past, this technique has been used in various fields such as consumer-watch, making future predictions and analyzing and interpreting large datasets for deeply embedded rules and associations. The qualitative details of Singapore tourists' foursquare check-ins, represented in a tabular form, is an example of a sequential database. Therefore, the Pattern-Growth method which uses Prefix-Span Algorithm is used in this study to obtain the Tourist Sequential Activity Patterns. Insights into tourist movement and activity patterns is deemed beneficial for the tourism sector in many ways, such as designing better travel packages for tourists, maximizing the tourist activity participation and meeting the tourist demands. This research proposes to adopt mobile social media data for effective capturing of tourist activity information in Singapore and utilizes advanced data mining techniques for extracting valuable insights into tourist behavior. The proposed methods and findings of the study have the potential to support tourism managers and policy makers in making better decisions in tourism destination management.
\end{abstract}


\section{Introduction}
With the rapid developments in information and communication technology (ICT) and in the era of globalization, the physical barriers of space and time have been eliminated with the inception of the Internet and World Wide Web. This has allowed business organizations to fully exploit the potential advantages offered by this new technology, allowing them to connect with their consumers online and offering them wide ranging products online. Further developments in ICT tools have also lead to the development of social media networks, blogs, RSS feeds, micro-blogs and wikis, which have allowed business organizations to directly engage with their consumers. Through these communication opportunities, many brands have connected with their consumers directly to improve their productivity, performance and profitability. This approach has also been adopted by the tourism industry. In recent times, the global tourism and hospitality sector has become highly competitive and therefore, tourists firms have adopted ICT  tools to engage with their consumers to design products according to their needs and requirements and to provide high quality products, at affordable prices. The developments in ICTs have led to the development of Location Based Social Networks (LBSN), which are commonly used by travelers in recent times. A LBSN is a social network that allows individuals to connect and share their location by sharing content including photographs, texts messages and videos. LBSN has the ability to track check-ins in real time, based on the location of the user. A fundamental benefit of using LBSNs to research on tourists' behavior is the accuracy and quality of data including texts, photos, videos and physical location coordinates. Based on the nature of the research, this study will aim at using LBSN and data mining techniques to investigate tourist activity pattern. It will also aim at proposing a new method for analyzing LBSNs by using sequential activity analysis and will focus on analyzing its practical relevance in terms of tourists' behavior, using Singapore as the case study. Finally, it will analyze the tourists' sequential activity patterns, which can be used by industry professionals and policy makers to utilize it to predict tourist behavior in terms of destination selection and can aid them in strategic planning and tourism marketing strategy as well as in new product and services development.  \par
Tourism is one of the leading contributors to the economic growth and development not only in Singapore but across the world \cite{Koz16}. Singapore is known for its richness in the attraction sites coupled with multicultural population and tropical climate. As a result, tourism has been the most active sector receiving thousands of tourists on an annual basis. According to the statistics from the Singapore Tourism Board, the tourism sector has realized a steady growth since September 2016 in terms of the international arrivals \cite{Koz16}. Subsequently, there has been an increase in visitation by 4\% last year, thus bringing an annual cumulative number of tourists that visited Singapore in 2017 to 8.5 million \cite{Bat15}.\par 
There are different factors which influence the tourists' behaviors and their choice of destinations \cite{Koz16}. In northern Asia and especially Singapore, the culture which includes dressing, language, art, music, religion and food is considered to be the most significant factor affecting tourism activities. Ideally, the more the cultural practices, it translates to shorter travel periods hence more destination visits and purchase of packaging tours by the inbound tourists. According to Reference \cite{Koz16}, the greater the cultural distance between a country and a destination, fewer tourists are likely to visit that particular destination. Based on convenience, variety and safety, Singapore emerged top of the ranks as the most preferred destination by tourists. This has been attributed to the government initiative to effect laws and regulations that ensure the security of the visitors in the destination of their choice \cite{Li2011, Wang2018, Wang2015}. \par

To study tourist activity patterns, we need to determine how close or similar a tourist activity pattern is relative to that of another tourist. An effective measure of the distance or similarity between the tourist activities should take into account many characteristics of tourist activities apart from spatial and temporal dimensions. Research has proven that tourists conduct different activities at certain times, therefore, when comparing individual activity patterns, differences in the attributes of these activities (e.g., type and purpose), should also be captured \cite{kim2010human}. The interdependency among these dimensions needs to be maintained at a distance or similarity measure (e.g., certain activities can take place only at certain places and/or at certain times). When comparing activity patterns, the distance measure should also be able to compare structural differences in tourist activities and their contextual variables (e.g., certain activities have to be performed before specific other activities)\cite{kim2010human}. Thus, the tourist activity patterns will unfold over time. It is evident from previous research that tourist behavior can be volatile and easily affected by a series of inter-twined factors \cite{bauder2015visitor,beeco2013gps,dejbakhsh2011cultural,le2015factors}. Thus, we attempt to find some regularity in the extracted patterns. Thereby contributing to the knowledge about tourist behavior related to the particular order of activities they prefer to choose. This might also be dependent on some extrinsic as well as intrinsic factors. For example tourist visiting a place might choose to take part in recreational activities like sight-seeing, and afterwards go for dining and after that choose to shop around the place. Studying these series of activities carried out by tourists on a large scale can give us an insight into tourist trends, choices and preferences which can assist in the understanding of tourist behavior. The knowledge gained can be applied for the strategic development of tour management and can lead to more sophisticated tour packages and travel itineraries.\par
In tourism research, Sequential Pattern Mining (SPM) has been extensively applied by researchers to try and understand the tourist behavior while visiting different destinations across the world. There are various methods that have been utilized in the past to understand tourist behavior \cite{Han13}. The most commonly used method is the vertical formatting method which entails the use of the Apriori algorithm in mining the tourist sequential activity pattern.This study will reveal various interesting patterns related to tourist activities, however due to the confined nature of this research, only the sequential activity patterns of tourists at diverse locations will be extracted and interpreted. The study will further seek to establish the most influential patterns on tourism activities at different venues in Singapore and their preferred timings. The analysis will be conducted on the data collected from Foursquare Check-ins through the Twitter API Streaming between October 2016 and October 2017 by using the Pattern-Growth SPM method. 
\textbf{Research Questions:}
\begin{enumerate}
	\item What are the most common tourist sequential activity patterns in Singapore?\par

	\item What are the effective methods of sequential activity analysis using the LBSN data in Singapore?\par

	\item What are the most interesting insights resulting from the tourist sequential activity pattern analysis in Singapore?
\end{enumerate}\par

\section{Literature Review}
Over the past three decades, there have been inconsistent and spatial tourist mobility patterns across the world and especially the Asia-Pacific region \cite{Bat15}. However, in the last decade, a steady but rapid growth of both inbound and outbound travel was noticed by the Pacific Asia Travel Association \cite{Hal12}. In the assessment of the tourist spatial patterns and flows in the early 1980s, it was established that these flows resulted due to the political and economic prosperity in the Asia-Pacific region. But owing to the rapid growth in the tourism sector, the researchers have become interested in the tourism flows in terms of nature, patterns and intensity.There are two approaches that influence the state of the tourism patterns and activities in any given region. The political-economic approach is often uni-directional and relies on economic situations to define the sequential patterns\cite{LiX08}. On the other end, the supply-demand interaction approach shapes tourism movement as well as consumption based on individual and collective preferences in a given tourist generation. For example, places with more tourism resources are poised to attract more tourists facing an upward shift on the frequency of the patterns.With the time and cost constraints, it creates intervening destinations thus providing avenues where tourists are able to make comparisons and select suitable destinations. Moreover, the flows and movement of tourists in the Asia-Pacific countries depend on other factors such as marketing effectiveness, promotional offers, destination attributes and demographic characteristics \cite{LiX08}.\par
Several countries have adopted various tools and methods to complete the initial surveys / research and comprehend their tourists better, by establishing tourists mobility patterns and flows.  Among the commonly used tools include the Country Potential Generation Index (CPGI) and the Gross Travel Propensity (GTP). The latter is used to evaluate the capability of a region or country to generate trips, taking into account a particular population. In essence it provides an estimate of the travel trips in a given region of the country.The Asia-Pacific region has gained attention from the world in terms of tourism owing to the economic and demographic development including less inbound travel restrictions \cite{Hal12}. Owing to the great potential of tourism growth and development, there is a need for Asia-Pacific countries to better comprehend tourist activity patterns and flows to enhance their management. The statistical report of the Pacific Asian Travel Association (PATA) exhibited different travel flows among the member countries meaning the tourist activity patterns are unique to a region. In 1995, the United States was the leading tourist destination followed by Canada, Hong Kong, Singapore, China, and Australia. However, the trends of the travel flows between 1995 and 2004 changed with China and Hong Kong topping the list of the leading destinations \cite{LiX08}. \par 

There is a substantial need to understand the tourists movement behavior with respect to their destination choices since it has a deep impact on tourism development process and marketing strategy. Many of the previous studies have been conducted in investigating tourist behavior with respect to destination by primarily emphasizing on spatial and temporal patterns of tourist \cite{zheng2017understanding,bauder2015visitor,kadar2014measuring}. There has also been an evaluation of most popular tourist trends and choices \cite{yang2013modeling,Yan17,Rua05}. The analysis of literature suggests that the tourist consumer behavior is the area of discussion of various researchers, academics and industry professionals because of its substantial importance. According to Reference \cite{Koz16}, consumer behavior is the widely researched area in tourism and marketing and therefore, it is frequently associated with tourist behavior or travel behavior. It is necessary to understand the individual tourist behavior to identify which factors influence their purchase intention and destination choice since they can be beneficial in influencing the tourism demand. The primary focus on tourist decision process offers a comprehensive, accurate and detailed analysis on their demands. Many of the researchers and industry professionals have focused on understanding tourist behavior to improve their overall marketing strategy, product development and overall quality of services offered. It is essential that travel agencies understand the tourists' destination preferences, since it allows them to develop new products and services by utilizing appropriate marketing strategies. It is also essential to understand the travel patterns of tourists to identify their behavior and purchase intent. It helps to identify their competitors and allows them to develop products and services based on different consumer segments, according to their needs and requirements. \par

In the past, there have been several design methods that were used to capture and analyze data on travel patterns and behaviors of tourists \cite{Yan17, Wang2002, Zhang2014}. Social media platforms, such as Flickr, have helped provide rich data sources in terms of the historical data of the tourists and their individual preferences. With millions and thousands of tourists visiting different places across the globe, the information from that platform assists in planning trips appropriately, especially to those unfamiliar cities. Ideally, the hosting service uses Geo-tagged photos to identify tourists' trajectories in a bid to explore topological spaces with the adoption of the motifs concept to unearth the tourists' mobility patterns. Modern tourists prefer to travel to different cities or places to spend their holiday, hence they require adequate information in terms of tourist trajectories and past trends in order to make substantial decisions. Flickr and Twitter are known to be the contemporary social media platforms that help to provide the most convenient tourist and travel recommendations to the users \cite{Zim17}. However, the privacy and the scalability issues have made the use of Flickr ineffective \cite{Sun2012, Sun2011, Sun2008}. Nonetheless, the travel recommendations can either be generic or personalized  \cite{Yan17}. The latter highlights individual preferences with regard to the matching of the locations during visits. On the other hand the generic recommendations follow a specific order that includes: trajectory identification > interesting locations > travel sequences > planning > activity recommendation. \par
Tourist trajectories comprise of sequences of landmarks with semantic, temporal, and spatial information  \cite{Kim17}. The use of the trajectory methods in understanding travel activity patterns requires the separation of the native tourists and the international tourists to better understand the flows. Flickr works by accumulating large collection of photos and storing the meta-data namely: size, time, and location. The latter is crucial in the analysis of the tourist activity patterns in a given region or place \cite{Agg13}. These methods collect the meta-data, analyze and provide recommendations accordingly to the users. The Geo-tagged photos help in partially capturing the travel information which can be used to construct travel trajectories and eventually unearth the tourist activity patterns for a given location. Consequently, the similarity matrix framework obtained after the construction of the respective trajectories help in grouping tourists accordingly. From the analysis of the travel semantic motif of the tourists, it was established that more tourists preferred the natural parks in the sequence as opposed to state buildings i.e. Central park > Brooklyn bridge > Rockefeller center. Different analysis using the trajectory framework from the meta-data obtained from the Flickr gets unique sequences depending on the tourist activity of that particular location \cite{Yan17}. Notably, tourists with similar interests in terms of travel preferences are normally grouped or clustered together generates the possibility of developing behavioral patterns and recommendations.\par 
Proper understanding of tourists' activities is important for tourism management to ensure they provide the best service, meet tourist expectation and also gain repeat visitation.There have been some attempts made in the past in studying the touristic flows and predicting tourist future destinations. One of the early contributions done in this respect was from Reference \cite{yang2013modeling} who focused on the decision making process of tourists and tried to model their next destination by using the nested logit model. This model assumed the utility maximization as the tourists demand and targeted on only one subsequent destination, unlike tourists, who can visit multiple subsequent destinations. Another research was done by Reference \cite{zheng2017understanding} who tried to model the tourist next destination through a survey group who collected data on tourists' intra-attraction spatial-temporal behavior and demographic characteristics using handheld GPS tracking devices and activity diary questionnaires. One of the latest research done in this category is by Reference \cite{vu2015exploring} who explored the travel behaviors of tourists in Hong Kong by using the data from Geo-tagged photos uploaded on Flickr. The tourists' movement trajectories were highlighted and patterns were drawn to indicate the most popular tourists destinations. The location preferences was also categorized with respect to 2 main groups i.e. Asian and Western. A similar study was carried out while exploring visitors activities in Hong Kong Parks \cite{vu2016exploring} and Temples \cite{yeung2016japanese}. The Twitter Streaming API has been used by Reference \cite{chua2016mapping} in which the Geo-tagged social media data is used to categorize tourists flow in Italy. This research used the Geo-tagged social media data from Twitter to characterize spatial, temporal and demographic features of tourists' flow in Cilento, Southern Italy. The study focused on 3 main areas which are: the tourists profiles, tourists travel patterns in the region, tourists attraction in the region and their popularity. The Geo-tagged photos on Flickr have been used in many other studies \cite{kadar2014measuring}. However, Flickr only gives the geo-coordinates of the photos without giving any information about the actual place, its type, category and user comments associated with it. What the tourist actually did at that particular location, which activities they were involved in, and how much time they spent at that particular activity cannot be assessed via the data obtained from Flickr. 
The past research has been beneficial in understanding tourists' behavior, however, little information is gained about tourist activity at a particular destination which is very important for tourism management and can assist in many ways. This research will fill the gap in the tourism literature by studying tourist activities in the sequential order, thereby providing rich information about tourist choices, preferences and decisions while visiting a particular destination.

\section{Methodology}
The Singapore dataset meets all the requirements to be regarded as the sequence database; essentially one tourist being involved in various activities with related time field. The Prefix-Span algorithm has been proven to outperform the Apriori algorithm including other emerging algorithms for SPM \cite{Han13}. Hence, the Pattern-Growth method which uses Prefix-Span algorithm will be used in this study.\par 
In terms of the execution time, the SPAM algorithm performs better than the Prefix-Span. The test performance evaluation was conducted using the `BMS Webview1' dataset which contained approximately 30,000 sequences with an average of 2.3 item sets in each of the sequences. The results of the evaluations are presented in the figure I below.\par
\begin{figure}[htbp]
\subfloat[]{\includegraphics[width=0.45\textwidth]{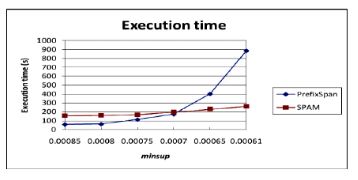}}\\
\subfloat[]{\includegraphics[width=0.45\textwidth]{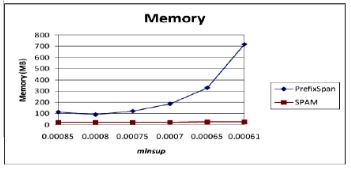}}
\caption{Evaluation of SPAM and Prefix-Span algorithm using the BMS dataset.}
\end{figure}

The Foursquare Check-ins formed the main dataset, which were collected via twitter streaming using the twitter developer API to extract all tweets of tourists visiting various destinations in Singapore. Using a specialized coded program, the raw dataset was filtered to obtain only the tweets with foursquare check-ins of the tourists. The data was collected for a period of between seven to nine months which included the category of venues to help figure out which activities the tourists engaged in at a particular time. Each check-in had the following attributes: check-in ID, user ID, time and geographic coordinate (latitude and longitude), category and subcategory of the check-in's location, i.e. the type of place where it occurred. Notably, the ideal dataset was collected for sample period of 4-6 months there were a total of 10,000 selected check-ins that were generated by 1057 number of tourists visiting singapore within the period of 4-6 months on an average each tourist generated 8-10 check-ins on Foursquare which was included in this study. 

\subsection{Prefix-Span Algorithm Development}

With such a convention, the expression of a sequence is unique. Next, we examine whether one can fix the order of item projection in the generation of a projected database. Intuitively, if one follows the order of the prefix of a sequence and projects only the suffix of a sequence, one can examine in an orderly manner all the possible sub-sequences and their associated projected database. Thus, we first introduce the concept of prefix and suffix. \\
\uline{Definition 1 (Prefix):} Suppose all the items within an element are listed alphabetically. Given a sequence a=<e1,e2,e3,$ \ldots $ en>(where each ei corresponds to a frequent element in S), a sequence B=<e1',e2'$ \ldots $ en'> is called a prefix of A if and only if 1) ei' = ei for (i<=m-1); 2) em' subset of em ; and 3) all the frequent items in (em-em') are alphabetically after those in em'.\\
\uline{Definition 2 (Suffix):}\ Given a sequence A=<e1,e2,..en> (where each ei corresponds to a frequent element in S). Let B=<e1e2..em-1em'> (m<=n) be the prefix of A. Sequence S=<em$"$ Em+1$ \ldots $ En> is called the suffix of A with regards to prefix B, denoted as C=A/B, where e$"$ =(em-em')$ \string^ $2 We also denote A=B.C Note, if B is not a sub-sequence of B, the suffix of \_ with regards to A is empty.\\
\uline{Definition 3 (Projected database):} Let A be a sequential pattern in a sequence database S. The \_-projected database, denoted as Sj, is the collection of suffixes of sequences in S with regards to prefix .To collect counts in projected databases, we follow the next step.\\
\uline{Definition 4 (Support count in projected database):} Let A be a sequential pattern in sequence database S, and B be a sequence with prefix A. The support count of B in A-projected database Sj, denoted as support Sj (B), is the number of sequences in Sj such that B is a subset of A. 

\subsection{The Pattern-Growth Method: Prefix-Span Algorithm}
\noindent The Pattern-Growth Method of mining sequential patterns involves the use of the Prefix-Span algorithm which can be executed in the following steps;
\begin{enumerate}
	\item Finding\ the length of the sequential patterns: The given database is scanned for all the frequent items of a given length. The sequences patterns ought to be of the same length.  \par

	\item Partitioning of the sequential patterns into subsets: The partitioning is done based on the attached prefix. \par

	\item Finding subsets of the sequence patterns: From the subsets of the sequential patterns,\ the\ projected\ database\ is constructed recursively.     
\end{enumerate}

The parameters involved are thus; S is the sequence database, $ \alpha $  is sequential pattern, l is the length of $ \alpha $ , SP is the alpha projected database. The major cost of the prefix-span is the generation of the projected database.  The process of the pattern-growth method involves a series of steps which requires input, method, and then output. The output in this case is the complete set of sequential patterns. Minimum support thresholds and confidence levels are required for the implementation of the algorithm.\\\\ 
\textbf{Method:}\\
\begin{itemize}
\item Scan the sequence database S to yield all the frequent items b where, b can be assembled to form sequential patterns or <b> may be appended together to $\alpha$ to form set of sequence patterns. 
\item Each of the frequent items form the projected database b is appended to form $\alpha$ which is the output.
\item In each of the given sequential patter,  the sequential pattern is constructed through the call of the prefix-scan ($\alpha$, l+1, SP).
\end{itemize}
The subsets can be formed by the resulting projected databases as shown in the Table 1 :\\\\

\begin{table}[htbp]
\label{tab:proDatabaseAndSeqPat}
\caption{Projected database and Sequential Patterns}
\begin{tabular}{|p{0.1\textwidth}|p{0.15\textwidth}|p{0.15\textwidth}|}
\hline
 \textbf{Prefix} & 
\textbf{Projected (postfix) database} & 
\textbf{Sequential Patterns} \\
\hline
a & ((abc) (ac) d (cf)),\newline  ((\_d) c(bc) (ae)),((\_b)\newline (df) cb), ((\_f) cbc) & (a), (aa), (ab), (a(bc)),\newline(a(bc)a), (aba), (abc), ((ab)),\newline((ab)c), ((ab)d), ((ab)f),\newline ((ab)dc), (ac), (aca), (acb),\newline(acc), (ad), (adc), (af) 
 \\
\hline
b & ((\_c)d(cf)), ((\_c) (ae),\newline ((df)cb), (c)  & (b), (ba), (bc), ((bc)), ((bc)a),\newline  (bd), (bdc), (bf)\\
\hline
c & ((ac)d(cf)), ((bc)(ae)), (b), (bc) & (c), (ca), (cb), (cc) \\
\hline
d & 
(cf) (c(bc)(ae)), ((\_f)cb)
&(d), (db), (dc), (dcb)\\
\hline
e & 
 (\_f)(ab)(df)cb),\newline ((af) cbc) & 
(e), (ea), (eab), (eac), (eacb),\newline  (eb), (ebc), (ec), (ecb), (ef),\newline (efb), (efc), (efcb)  \\
\hline
f & 
 ((ab) (df) cb), (cbc) & 
 (f), (fb), (fbc), (fc), (fcb)  \\
\hline
\end{tabular}
\end{table}

In a given sequence dataset, in this case, the Singapore Tourist Database, the Pattern-Growth method using the Prefix-Span algorithm helps in identifying all the frequent sequence patterns in the dataset. There are two parameters which are executed at the start of the algorithm, namely; minimum support and maximum prefix \cite{Sar15}. The latter helps by providing the length of the sequence which is quite crucial while analyzing large databases. On the other end, the minimum support parameter is obtained by dividing the pattern with the number of sequences in the dataset. Using the projected database, the Prefix-Span algorithm first finds out the lengths of the sequential patterns then uses the projected database to yield the patterns.

\begin{table}[htbp]
\centering
\caption{An Example of a sequence database}

\begin{tabular}{|l|l|}
\hline
 \textbf{ID} & 
\textbf{Sequence}  \\
\hline
S1 &  (1),(2),(1 2),(3),(1 3),(4 5),(6) \\
\hline
S2 & (3 4),(3),(2 3),(1 4)  \\
\hline
S3& (4 5),(2),(2 3 4),(3),(1)\\
\hline
S4& 
(4),(5),(1 6),(3),(2),(7),(1)\\
\hline
\end{tabular}%

\end{table}

The Table 2 represents an example of the sequence database which contains four sequential patterns having items sets. For instance, the first sequence contains seven item sets of frequent patterns \cite{Sar15, Sun2011b}. Similarly, the analysis of the Venue Category and Check-in Time shall yield the subsequent patterns from the selected sequence of the Singapore Database. The application of the Prefix-Span entails the scanning of the sequence database to obtain the frequent items \cite{han2004sequential, Hu2006, Li2017, Peng2018}. Thereafter, the frequent items are appended together to form sequential patterns which is then used in the construction of the projected database.\par  
The prefix span algorithm shall be implemented in java pseudo code although it can also be run using python and R programming languages. 

\section{Results and Findings}
The findings of the SPM using the Pattern-Growth method are summarized in five primary cases each exhibiting different activity patterns altogether. It provides an indication that tourists have different tastes and preferences while visiting a particular region in the world. The results provide adequate information to the tourists' management authorities in Singapore to help them devise better strategies in terms of tourists' visits.
\begin{figure}[htbp]
\includegraphics[width=0.5\textwidth]{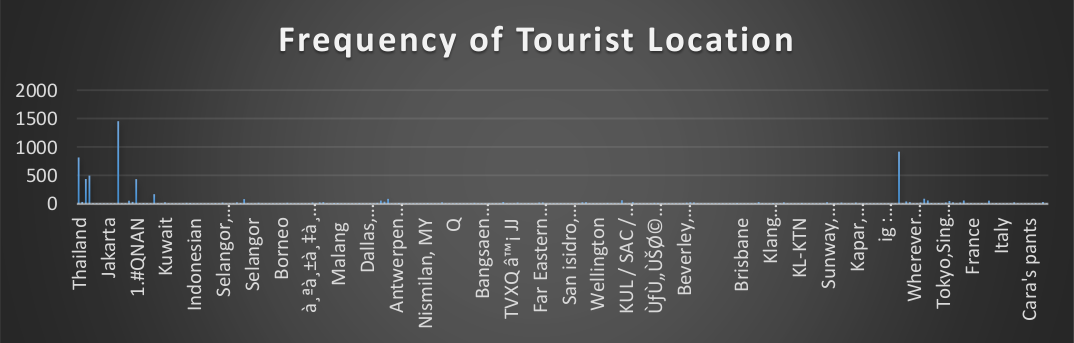}
\caption{Frequency of the Tourist Location}
\end{figure}
From figure 2, it is evident that most of the tourists came from Thailand, Kuwait, Jakarta Malaysia, and Indonesia.

The exploratory data analysis revealed that there were more female tourists than male tourists visiting different venues in the country. Ideally, there were 3830 female, 3577 male, and 207 tourists did not wish to declare their gender. From the analysis, majority of the female tourists came from Malaysia, NaN, Thailand, Peru, and BBK. On the other end, majority of the male tourists came from Japan, NaN, Kuwait, Selangor, Thailand, and Indiana.
\begin{figure}[htbp]
\includegraphics[width=0.5\textwidth]{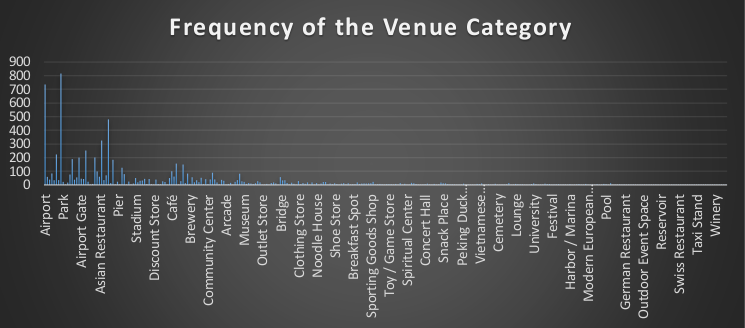}
\caption{Frequency of the Venue Category}
\end{figure}
Most of the visited Venue Categories include: Airport, Park, Airport-Gate, Asian Restaurant, Pier, and Stadium. The scenic lookout and the airport was visited more frequently as compared to other activities. We already know that Airport and all the Airport related activities signal the international tourists status, therefore, these will be ignored when generating activity patterns in the SPM phase.\par
Using the Pattern-Growth method and the Prefix-Span algorithm the following sequence patterns were obtained. The Apache software which runs on java was used to implement the algorithm following the systematic steps to yield the tourist sequential activity patterns.

\begin{table}[htbp]
\caption{Case 1: Tourists' Sequential Activity Patterns in the Morning (7am-2pm)}
\resizebox{\linewidth}{!}{%
\begin{tabular}{|l|l|l|l|}
\hline
\textbf{Activity Sequence 1 > 2 > 3 } & 
 \textbf{Frequency} & 
\textbf{Support} & 
\textbf{Confidence} \\
\hline
\textbf{Travelling > Religious > Dining} & 
620 & 
 0.015713 & 
 0.078125 \\
\hline
\textbf{Shopping > Recreation > Entertainment} & 
 620 & 
 0.136509 & 
 0.030576 \\
\hline
\textbf{Hiking > Outdoor > Refreshments} & 
 385 & 
 0.042229 & 
 0.098837 \\
\hline
\textbf{Entertainment > Religious > Shopping} & 
 289 & 
 0.283575 & 
 0.030303 \\
\hline
\textbf{Shopping > Hiking > Dining} & 
 365 & 
 0.044685 & 
0.192308 \\
\hline
\textbf{Nature > Refreshments > Dining} & 
 265 & 
0.136509 & 
 0.008993 \\
\hline
\textbf{Archives > Shopping > Nature} & 
423 & 
 0.006629 & 
 0.185185 \\
\hline
\textbf{Dining > Travelling > Nature} & 
 313 & 
 0.005892 & 
 0.208333 \\
\hline
\textbf{Hiking > Nature > Shopping} & 
313 & 
 0.136509 & 
0.008993 \\
\hline
\textbf{
Dining > Nature > Nature} & 
 287 & 
 0.083231 & 
 0.014749 \\
\hline
\textbf{Sport > Dining > Nature} & 
 465 & 
 0.012521 & 
 0.098039 \\
\hline
\textbf{Nature-walk  > Archives > Nature} & 
 550 & 
 0.03781 & 
 0.032468 \\
\hline
\textbf{Religious > Nature  > Shopping} & 
 425 & 
 0.009084 & 
 0.135135 \\
\hline
\textbf{Entertainment > Nature > Refreshments} & 
 414 & 
 0.098453 & 
 0.014963 \\
\hline
\textbf{Dining > Hiking > Shopping} & 
 611 & 
 0.017677 & 
 0.083333 \\
\hline
\end{tabular}%
}
\end{table}

Although, primarily tourists varied in their times of involvement in various activities, there were notable activities such as visiting the park, scenic look-out, outdoor sculpture, visiting the garden and border crossing that were commonly done in the morning hours. Hiking, nature walks and religious activities dominates the morning hours of the tourist schedules based on the patterns, as evident in case 1. As compared to other time periods, afternoon and evening, people prefer to visit different restaurants to eat in the morning hours. Based on the figure above several tourists share most of the sequential activity patterns.

\begin{table}[htbp]
\caption{Case 2: Tourists' Sequential Activity Patterns in the Afternoon (2pm-12am)}
\resizebox{\linewidth}{!}{%
\begin{tabular}{|l|l|l|l|}
\hline
 \textbf{Activity Sequence 1 > 2 > 3} & 
 \textbf{Frequency} & 
\textbf{Support} & 
 \textbf{Confidence} \\
\hline
\textbf{Dining > Walking > Shopping} & 
 715 & 
0.050011 & 
 0.088235 \\
\hline
\textbf{Field > Exhibition > Shopping }& 
715 & 
 0.126917 & 
 0.008278 \\
\hline
\textbf{Field > Dining > Shopping} & 
 385 & 
 0.007354 & 
 0.142857 \\
\hline
\textbf{Shopping > Gaming > Walking} & 
270 & 
 0.075856 & 
 0.033241 \\
\hline
\textbf{Dining > Walking > Shopping} & 
365 & 
 0.054003 & 
 0.046693 \\
\hline
\textbf{Nature > Nature > Walking} & 
 444 & 
 0.126917 & 
 0.013245 \\
\hline
\textbf{Travelling > Dining > Walking} & 
 323 & 
 0.007144 & 
 0.235294 \\
\hline
\textbf{Shopping > Refreshments > Refreshments} & 
 313 & 
 0.024795 & 
 0.042373 \\
\hline
\textbf{Nature > Dining > Walking} & 
513 & 
0.050011 & 
0.021008 \\
\hline
\textbf{Shopping > Field > Nature} & 
 587 & 
0.093717 & 
0.013453 \\
\hline
\textbf{Dining > Field > Hiking} & 
265 & 
0.014079 & 
0.089552 \\
\hline
\textbf{Archives  > Travelling > Resting} & 
240 & 
0.075856 & 
0.01662 \\
\hline
\textbf{Scenery > Dining > Entertainment} & 
125 & 
0.037403 & 
0.033708 \\
\hline
\textbf{Dining > Archives > Walking} & 
114 & 
0.007354 & 
0.171429 \\
\hline
\textbf{Archives > Dining > Field} & 
211 & 
0.126917 & 
0.034768 \\
\hline
\end{tabular}%
}
\end{table}


The case 2 features those set of sequence which were generated within the time range of mid-afternoon 2pm till mid-night 12am. The activity patterns clearly highlight the trend of tourists during the later part of the day. While a large number of tourists preferred to do some nature and field walks, dining formed the major activity at this time. Thus, they have plenty of options to choose from while dining out. This also symbolizes the rich Asian culture that predominates the world due to it's authentic and aromatic variety of food.\ \  
\\

\begin{table}[htbp]
\caption{Case 3:Most predominant Sequential Activity Patterns among Tourists}
\resizebox{\linewidth}{!}{%
\begin{tabular}{|l|l|l|l|}
\hline
 \textbf{Activity Sequence 1 > 2 > 3} & 
 \textbf{Frequency} & 
\textbf{Support} & 
 \textbf{Confidence} \\
\hline
\textbf{Hiking > Dining > Shopping} & 
 1200 & 
0.01458
 & 
 0.235294 \\
\hline
\textbf{Shopping > Field > Nature}& 
1300 & 
 0.14494
 & 
 0.031558
 \\
\hline
\textbf{Field > Dining > Shopping} & 
 385 & 
 0.007354 & 
 0.142857 \\
\hline
\textbf{Shopping > Nature > Dining} & 
785& 
 0.075856 & 
 0.033241 \\
\hline
\textbf{Dining > Walking > Shopping} & 
365 & 
 0.042882
 & 
 0.106667
 \\
\hline
\textbf{Dining > Shopping > Outdoor
} & 
 870 & 
 0.311607
 & 
 0.03211
 \\
\hline
\textbf{Heritage Trail > Nature > Outdoor} & 
 965 & 
 0.047742
 & 
 0.209581
 \\
\hline
\textbf{Dining > Nature > Religious
} & 
 344 & 
 0.14494
 & 
 0.009862
 \\
\hline
\textbf{Shopping > Religious > Heritage Trail
} & 
523 & 
0.007433
 & 
0.192308
\\
\hline
\textbf{Religious > Hiking > Shopping
} & 
 613
 & 
0.091481
 & 
0.015625
 \\
\hline
\textbf{Entertainment > Walking > Dining
} & 
723 & 
0.014294
 & 
0.1 \\
\hline
\textbf{Shopping > Refreshments > Hiking} & 
287
 & 
0.043453
 & 
0.026316
\\
\hline
\textbf{Dining > Archives > Outdoor
} & 
965 & 
0.011435
 & 
0.1\\
\hline
\textbf{Shopping > Nature > Dining
} & 
965
 & 
0.311607
 & 
0.011009
 \\
\hline
\end{tabular}%
}
\end{table}

Both male and female tourists were largely involved in Shopping, Nature walk, and Dining. The tourists prefer to visit variety of hotels and restaurants to explore different types of food owing to the rich Asian culture. The activity pattern  Shopping > Field > Nature had the highest frequency indicating that most of the tourists spent substantial hours of their daytime involving in recreational and other outdoor activities.


\begin{table}[htbp]
\caption{Case 4: Sequential Activity Patterns of Female Tourists}
\resizebox{\linewidth}{!}{%
\begin{tabular}{|l|l|l|l|}
\hline
\textbf{Activity Sequence 1 \textgreater{} 2 \textgreater{} 3 } &
\textbf{Frequency} & \textbf{Support} &
\textbf{Confidence}\\ \hline
\textbf{Shopping \textgreater{} Travel \textgreater{} Nature } & 1700 &
0.010649 & 0.21875\\ \hline
\textbf{Dining \textgreater{} Travel \textgreater{} Religious} & 1480 &
0.115474 & 0.027378\\ \hline
\textbf{Dining \textgreater{} Refreshment \textgreater{} Archives} &
1650 & 0.034276 & 0.092233\\ \hline
\textbf{Hiking \textgreater{} Nature \textgreater{} Leisure } & 1400 &
0.255907 & 0.004551\\ \hline
\textbf{Refreshments \textgreater{} Dining \textgreater{} Arcade} & 1365
& 0.004659 & 0.25\\ \hline
\textbf{Dining \textgreater{} Shopping \textgreater{} Nature} & 1278 &
0.255907 & 0.030559\\ \hline
\textbf{Dining \textgreater{} Archives \textgreater{} Nature} & 1415 &
0.037604 & 0.207965\\ \hline
\textbf{Hiking \textgreater{} Nature \textgreater{} Entertainment} &
1498 & 0.255907 & 0.011704\\ \hline
\textbf{Field \textgreater{} Nature \textgreater{} Outdoor} & 1370 &
0.009151 & 0.327273\\ \hline
\textbf{Travel \textgreater{} Dining \textgreater{} Nature } & 1589 &
0.115474 & 0.063401\\ \hline
\textbf{Nature \textgreater{} Shopping \textgreater{} Outdoor} & 1220 &
0.083195 & 0.088\\ \hline
\textbf{Archives \textgreater{} Outdoor \textgreater{} Sporting} & 1120
& 0.255907 & 0.005202\\ 
\hline
\end{tabular}%
}
 \end{table}

It is clear from the above activity patterns, case 4, that most of the female tourists preferred to engage in shopping followed by nature walk, while others were involved in outdoor and sporting activities. This could also be attributed to the varying age groups of the female tourists, that were included in the dataset. 
\\


\begin{table}[htbp]
\caption{Case 5:Sequential Activity Patterns of Male Tourists}
\resizebox{\linewidth}{!}{%
\begin{tabular}{|l|l|l|l|}
\hline
\textbf{Activity Sequence 1 \textgreater{} 2 \textgreater{} 3 } &
\textbf{Frequency} & \textbf{Support} &
\textbf{Confidence}\\ \hline
\textbf{Shopping \textgreater{} Travel \textgreater{} Nature} & 1200 &
0.010318 & 0.209677\\ \hline
\textbf{Travel \textgreater{} Archives \textgreater{} Entertainment} &
1500 & 0.115327 & 0.027417\\ \hline
\textbf{Hiking \textgreater{} Entertainment \textgreater{} Gaming} &
1550 & 0.034948 & 0.090476\\ \hline
\textbf{Dining \textgreater{} Religious \textgreater{} Nature } & 1400 &
0.25778 & 0.027114\\ \hline
\textbf{Archives \textgreater{} Refreshments \textgreater{} Nature} &
1265 & 0.03578 & 0.195349\\ \hline
\textbf{Dining \textgreater{} Recreation \textgreater{} Nature } & 1178
& 0.25778 & 0.01162\\ \hline
\textbf{Nature \textgreater{} Dining \textgreater{} Market} & 1315 &
0.008654 & 0.346154\\ \hline
\textbf{Dining \textgreater{} Travel \textgreater{} Field} & 1498 &
0.009819 & 0.135593\\ \hline
\textbf{Nature \textgreater{} Outdoor \textgreater{} Religious} & 1370 &
0.08221 & 0.016194\\ \hline
\textbf{Field \textgreater{} Travel \textgreater{} Nature } & 1189 &
0.115327 & 0.063492\\ \hline
\textbf{Nature \textgreater{} Sporting \textgreater{} Gaming} & 1120 &
0.08221 & 0.089069\\ \hline
\textbf{Nature \textgreater{} Outdoor \textgreater{} Dining} & 1120 &
0.25778 & 0.008393\\ \hline
\end{tabular}%
}
 \end{table}

It can be drawn from the above cases that support threshold and confidence contributed significantly in the generation of the sequential patterns. Thus it can be concluded that (Travel > Archives > Entertainment) and (Hiking > Entertainment > Gaming), in case 5, had the highest frequency, implying that male tourists preferred entertainment activities during their trip.
Essentially, the study attempted to investigate the times in which the tourists checked in at their respective destinations. Notably, there were places where tourists checked in more than once such as the shopping malls, metro stations, hotels, scenic lookouts and food courts. For this research, a typical day begun at 07:00 hrs. and ended at 23:00 hrs. Hence, the activities done at the beginning of the day were placed at the top of the sequence patterns followed by other activities simultaneously.  

\section{Conclusion}
In every economy, tourists form the largest contributor to its growth and development. In regards, there is need to further understand all the dynamics to be able to improve on the service delivery to both international and domestic tourists. SPM is one of the most important and extensively used data mining technique that is applied in the tourism sector in order to understand the activity patterns of the visitors in various destinations. This study involved the use of the Pattern-Growth method to facilitate the SPM process into the formulation of the activity patterns. Although tourists of both genders shared some of the activity patterns, there were patterns that were independent of each other. The tourism management of Singapore will be able to use the insights of the most interesting patterns to understand the appropriate check-in times for the various activities hence will be able to prepare more tailored tour package as well as improve and customize the delivery of their travel services.  
\section*{Acknowledgment}
The authors gratefully acknowledge the valuable direction and support of Professor Hua Wang, Centre for Applied Informatics, College of Engineering \& Science and Research Fellow Huy Quan in the research and findings of this paper. Due to  the numerous perspective comments on the several drafts of the manuscript and signifying gaps in the knowledge, the author was able to compile and write this paper successfully. 
\IEEEtriggeratref{27}
\bibliographystyle{unsrt}

\end{document}